\RequirePackage{fix-cm}
\documentclass[smallextended]{svjour3}       
\smartqed  
\usepackage{graphicx}
\usepackage{cite}
\usepackage{balance}
\usepackage{amsmath,amssymb,amsfonts}

\usepackage{algorithmic}
\usepackage{textcomp}
\usepackage{soul}
\usepackage{xcolor}
\usepackage{multirow}
\usepackage[ruled,vlined]{algorithm2e}
\usepackage{url}

\newcommand{\tool}{{\it DRLinter}}

\begin{document}

\title{Faults in Deep Reinforcement Learning Programs: A Taxonomy and A 
Detection Approach 
}
\subtitle{}


\author{Amin Nikanjam \and Mohammad Mehdi Morovati \and Foutse Khomh \and Houssem Ben Braiek}


\institute{Amin Nikanjam \and
           Mohammad Mehdi Morovati \and
           Foutse Khomh \and
           Houssem Ben Braiek \at
              SWAT Lab., Polytechnique Montréal, Montréal, Canada\\
              \email{\{amin.nikanjam,mehdi.morovati,foutse.khomh,houssem.ben-braiek\}@polymtl.ca}           
}

\date{Received: date / Accepted: date}

\maketitle

\begin{abstract}
A growing demand is witnessed in both industry and academia for employing Deep Learning (DL) in various domains to solve real-world problems. Deep Reinforcement Learning (DRL) is the application of DL in the domain of Reinforcement Learning (RL). Like any software system, DRL applications can fail because of faults in their programs. In this paper, we present the first attempt to categorize faults occurring in DRL programs. We manually analyzed 761 artifacts of DRL programs (from Stack Overflow posts and GitHub issues) developed using well-known DRL frameworks (OpenAI Gym, Dopamine, Keras-rl, Tensorforce) and identified faults reported by developers/users. We labeled and taxonomized the identified faults through several rounds of discussions. The resulting taxonomy is validated using an online survey with 19 developers/researchers. To allow for the automatic detection of faults in DRL programs, we have defined a meta-model of DRL programs and developed \tool{}, a model-based fault detection approach that leverages static analysis and graph transformations. The execution flow of \tool{} consists in parsing a DRL program to generate a model conforming to our meta-model and applying detection rules on the model to identify faults occurrences. The effectiveness of \tool{} is evaluated using 21 synthetic and real faulty DRL programs. For synthetic samples, we injected faults observed in the analyzed artifacts from Stack Overflow and GitHub. The results show that \tool{} can successfully detect faults in both synthesized and real-world examples with a recall of 75\% and a precision of 100\%.

\end{abstract}

\section{Introduction}
\label{intro}
\begin{figure*}[t]
\begin{center}
\includegraphics[width=0.8\linewidth]{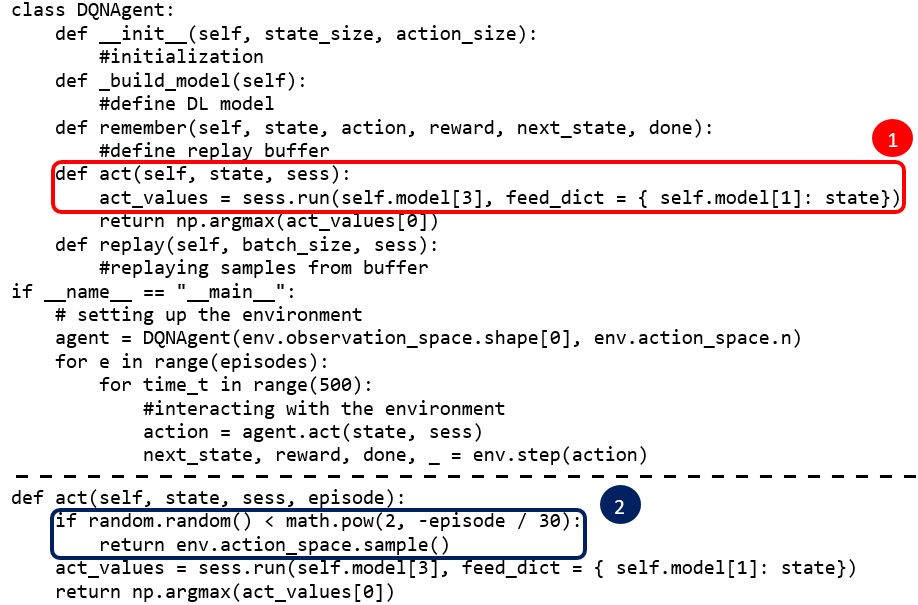}
\caption{Example of \textit{“Missing exploration”} fault from SO\#47750291.}
\label{fig:snippet1}
\end{center}
\end{figure*}

Applications of Deep Learning (DL) are growing in a variety of domains in both academia and industry. We are now seeing DL-based technologies being implemented in critical systems such as autonomous driving cars and medical diagnosis systems. Deep Reinforcement Learning (DRL) which is the application of DL in Reinforcement Learning (RL) is an active field of Machine Learning (ML), that exploits the capabilities of DL architectures to address previously unsolvable problems in RL. The importance of DRL is that it allows agents to infer decisions directly from unstructured input data (e.g., every pixel in the screen of a video game). DRL algorithms can deal with very  large  input  spaces  (in  contrast  to  traditional  RL)  and  indicate actions that optimize the reward (e.g., maximizing the game score). DRL has been widely applied to games \cite{schrittwieser2020mastering,silver2016mastering}, robotics \cite{levine2016end,akkaya2019solving}, healthcare \cite{yu2019reinforcement}, finance \cite{fischer2018reinforcement}, autonomous driving \cite{sallab2017deep} and navigation of high-altitude balloons \cite{bellemare2020autonomous}. While DRL-based systems are employed in various domains, their reliability is still a major source of concern. Compared to traditional software systems, the notion of faults in DRL-based software systems is more complex since: 1) they use complex DL architectures, 2) they employ RL algorithms that can be complex due to difficult hyperparameter tuning \cite{zhang2021importance} and sequential decision making, i.e., each agent’s decision can affect its future actions, and 3) DRL programs are interacting with non-deterministic environments.\\
In this paper, we aim to investigate the types of real faults occurring in DRL programs to construct a taxonomy of such faults and then, propose a model-based detection approach to find those faults in DRL programs. We study faults that could be encountered when implementing a DRL algorithm, i.e., a RL algorithm that benefits from Deep Neural Networks (DNN). Although other researchers have worked on categories of faults in DL programs \cite{DLtaxo2020, islam2019comprehensive} focusing basically on faults occurring in building and training a DNN, to the best of our knowledge, this paper is the first research work to study and categorize (in the form of a taxonomy) the types of faults that occur in DRL programs. Such taxonomy would help developers and researchers to improve their understanding of the root cause and symptoms of DRL faults. This would allow them to prevent common faults during the development of DRL systems, prepare test cases, and improve DRL frameworks for facilitating development and testing.\\
As a motivating example, Figure \ref{fig:snippet1} shows a faulty example extracted from Stack Overflow post \#47750291. In this example, the developer has employed  Deep Q-Network (DQN), a well-known DRL algorithm to solve the CartPole problem, a widely used benchmark problem in RL. The symptom is expressed as bad performance in terms of low reward by the developer. Actually, the agent got stuck at a suboptimal reward level without further improvement. However, lack of enough exploration was determined to be the root cause in the accepted answer. The developer forgot to consider an exploration strategy for the agent, so the agent always selects the action based on the output of Q-network (a DNN). In this way, the agent fails to perform random actions to gather information from the environment. Environment exploration is necessary for the success of RL agents and so this sample is identified as “Missing exploration” in our study. The location of fault and recommended modification is indicated in the code by (1) and (2), respectively, according to the accepted answer of the post.
\\
Our methodology for building the taxonomy consists in the manual analysis of faulty software artifacts. First, we have mined software repositories and Q\&A forums, i.e., GitHub\footnote{https://www.github.com} and Stack Overflow\footnote{https://stackoverflow.com} (SO), to find relevant artifacts. In these artifacts, developers/researchers discussed and--or fixed issues that occurred while they were using popular DRL frameworks. We have manually analyzed the artifacts and identified 761 issues. Next, we have categorized the relevant issues through a multi-round labelling process. 
At the end, 11 distinctive types of faults have been obtained that contain 27 faulty artifacts. We have validated the obtained taxonomy through a survey with 19 participants who have various backgrounds and levels of expertise. Then, we present an approach to detect such faults in DRL programs using static analysis which is called \tool{}. In this first attempt, we consider DQN which is a well-known DRL algorithm. A model-based approach using graph transformations is proposed to detect categorized types of faults occurring in DRL programs. A meta-model for DRL programs is proposed as a type graph, capturing the base skeleton and fundamental properties of DRL programs independent of available development frameworks. Considering the proposed meta-model, we specify for each fault type, a graph transformation rule to detect fault occurrences. DRL code is analyzed to extract its model conforming to the meta-model. Finally, a checking process is performed to verify models of DRL programs using graph transformations. We have successfully evaluated our approach by finding faults in 15 synthetic and 6 real faulty DRL programs. The results show that \tool{} effectively detects bugs in faulty codes. Briefly, this paper makes the following contributions:
\begin{itemize}
\item We propose an initial taxonomy of faults in DRL programs;
\item We describe 11 common errors of DRL programs according to real faulty samples;
\item We propose a model-based fault detection approach for DRL programs, using meta-modelling and graph transformation rules. 
\item We provide a concrete implementation of the approach as a tool. 
\end{itemize}
The rest of the paper is organized as follows. In Section \ref{drl}, we briefly review DRL and its approaches. We present a full description of our proposed taxonomy and the methodology followed to construct and validate it in Section \ref{taxonomy}. In Section \ref{modeling}, the proposed model-based fault detection approach is reported including the meta-model, the graph transformation rules, and the implementation details of \tool{}. Evaluation results of \tool{} are reported in Section \ref{exp}. A discussion with mentioning threats to validity is presented in Section \ref{validity}. Related works are reviewed in Section \ref{relatedwork} and finally, we conclude the paper in Section \ref{conclusion}.

\section{Deep Reinforcement Learning}
\label{drl}
\begin{figure}
\begin{center}
\includegraphics[width=0.5\linewidth]{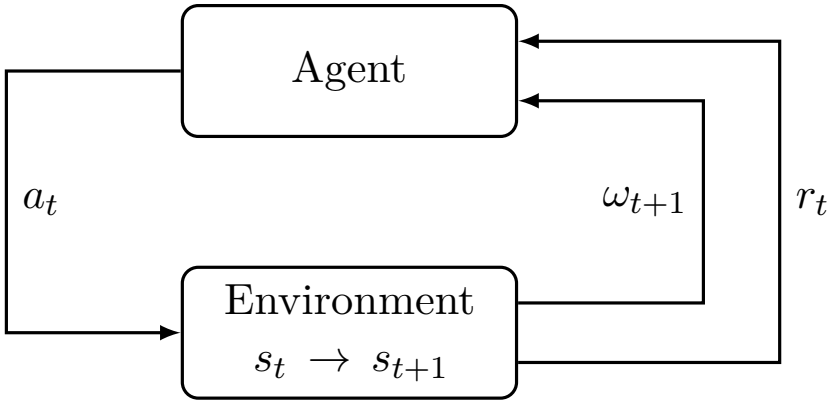}
\caption{Agent interacting with its environment \cite{LavetDRL2018}.}
\label{fig:agent-E}
\end{center}
\end{figure}

ML is classified into three main branches: Supervised Learning, Unsupervised Learning and Reinforcement Learning \cite{morales2019deep}. Supervised learning is mainly about inferring a classification/regression from labeled training data. The main goal of unsupervised learning is to draw inference from unlabeled input data. In RL, an agent interacts with an environment and the task consists in learning how to perform sequences of actions in the environment to maximize cumulative returning rewards. The agent aims to learn good behavior; meaning that it modifies its behaviour incrementally or attempts new ones. Moreover, agent uses trial-and-error experience, i.e., frequent interactions with the environment and information collection \cite{LavetDRL2018}. In other words, RL basically aims to handle the automatic learning of optimal decisions over time \cite{lapan2018deep,morales2019deep}.\\
Formally, the RL problem is formulated as a discrete-time stochastic control process in the following way: at each time step $t$, the agent has to select and perform an action $a_t$ from the set of actions $A$. Upon taking the action, (1) the agent is rewarded by $r_t \in R$ where $R$ is the set of rewards, (2) the state of environment is changed to $s_{t+1} \in S$ where $S$ is the set of states, and (3) the agent perceives the next observation of $\omega_{t+1}$ from the set of observations $\Omega$. Figure~\ref{fig:agent-E} illustrates such agent-environment interaction. An RL agent is defined to find a policy $\pi$ from the set of all policies $\Pi$ that maximizes the expected cumulative reward (a discount factor $ \gamma \in [0,1]$ applies to the future rewards) \cite{LavetDRL2018}. 
\\
Recently, researchers have successfully integrated DL methods in RL to solve some challenging sequential decision-making problems \cite{Goodfellow-et-al-2016}. This combination of RL and DL is known as deepRL or DRL. DRL benefits from the advantages of DL in learning multiple levels of representation among data to handle large state-action spaces with low prior knowledge. For example, a DRL agent has successfully learned how to play video games from raw visual perceptual inputs including thousands of pixels \cite{mnih2015human}. DRL algorithms, unlike traditional RL, are capable of dealing with very large input spaces, and indicating actions that optimize the reward (e.g., maximizing the game score). As a consequence, imitating some human-level problem solving capabilities becomes possible \cite{gandhi2017learning, moravvcik2017deepstack}. However, DRL is considered as a subfield of DL-based techniques, so it is not yet as popular as DL in general.
\subsection{Value-based approaches}
The value-based algorithms in RL aim to build a value function, which subsequently makes it possible to define a policy. The value function for a state is defined as the total amount of discounted reward that an agent expect to accumulate over the future, starting from that state. Q-learning algorithm \cite{watkins1992q} is the simplest and most popular value-based algorithm. In the basic version of Q-learning, a lookup table of Q-values, $Q(s, a)$, with one entry for every state-action pair is used to approximate the value function. To learn the optimal Q-value function, the Q-learning algorithm uses an incremental approach by updating Q-values with an update rule after taking an action.\\ 
The idea of \textit{value-based deep reinforcement learning} is approximating the value function by a DNN. Mnih et al. \cite{mnih2015human} introduced the DQN algorithm that obtained human-level performance in an online playing of ATARI games. They have defined the state as the stack of four consecutive frames and actions as various joystick positions. The deep network consists of multiple convolutional and fully-connected layers. DQN uses two heuristics to address its instabilities: 1) using target network: instantiating fitted Q-network for some iterations and applying network’s parameters update only periodically, and 2) replay memory (buffer): keeping all information of several previous steps and replaying them (as mini-batch) to reduce variance.

\subsection{Policy gradient approaches}
Policy gradient methods maximize a performance objective (typically the expected cumulative reward) by discovering a good policy. Basically, the policy function is directly approximated by a DNN meaning that the network output would be (probability of) actions instead of action values (say estimated rewards). It is acknowledged that policy-based approaches converge and train much faster specially for problems with high-dimensional or continuous action spaces \cite{agostinelli2018reinforcement}. The direct representation of a policy to extend DQN algorithms for addressing  the restriction of discrete actions was introduced by Deep Deterministic Policy Gradient (DDPG) \cite{lillicrap2015continuous}. This algorithm updates the policy in the direction of the gradient of Q which is a computationally efficient idea.\\
Another approach is using an actor-critic architecture which benefits from two neural network function approximators: an actor and a critic. The actor denote the policy and the critic is estimating a value function (e.g., the Q-value function). Asynchronous advantage actor-critic (A3C) algorithm \cite{mnih2016asynchronous} can employ both feed-forward and recurrent neural approximators to learn tasks in continuous action spaces, working both on 2D and 3D games.
\subsection{Review of frameworks for developing DRL programs}
In this study, we consider four popular frameworks for developing DRL programs: OpenAI Gym, Dopamine, keras-RL, and Tensorforce. It should be noted that we only study currently active frameworks over GitHub. We consider a framework as active if its repository is supported to resolve issues which are submitted to that repository. To compare these frameworks, we have gathered some information about them from their GitHub repositories. Table \ref{tbl:repo} presents detailed information about the studied frameworks. The number of stars of a repository is considered as one of the most important metrics for identifying repository popularity \cite{borges2016understanding}. The number of stars identifies the number of users who used and find the repository interesting. \textit{OpenAI Gym}, the most popular repository in our list, is a toolkit that provides a set of standardized environments for developing RL algorithms \cite{openaigym}. It is supported by \textit{OpenAI} and was first released in 2016. 
Dopamine is another popular repository. It is a framework for prototyping RL algorithms which is developed by \textit{Google} and released in 2018 \cite{castro18dopamine}.
Keras-rl is also a framework providing RL algorithms for \textit{keras}. It was launched in 2016 \cite{plappert2016kerasrl}. Last but not least, Tensorforce is a \textit{TensorFlow} library for applied RL algorithms \cite{tensorforce}. 
\begin{table}
\caption{Detailed information of selected frameworks.}
\label{tbl:repo}
\centering
\begin{tabular}{l r r r r} 
\hline
\textbf{\textit{Project Name}} & \textbf{\textit{stars}} &\textbf{\textit{commits}} & \textbf{\textit{issues}}& \textbf{\textit{contributors}} \\ 
\hline
OpenAI Gym   & $22k$  & 1,217  & 1,179  & 248 \\
Dopamine     & $9.1k$ & 197  & 118  & 8 \\
Keras-rl     & $4.8k$ & 308  & 214  & 39 \\
Tensorforce  & $2.7k$ & 1,979  & 512  & 60 \\
\hline
\end{tabular}
\end{table}

\section{The Taxonomy of Faults in DRL programs}
\label{taxonomy}
We have prepared a replication package that includes the materials used in this study and  anonymized data collected during our survey \cite{replication}.
\subsection{Methodology}
\label{methodology}
\begin{figure*}
    \centering
    \includegraphics[width=\textwidth]{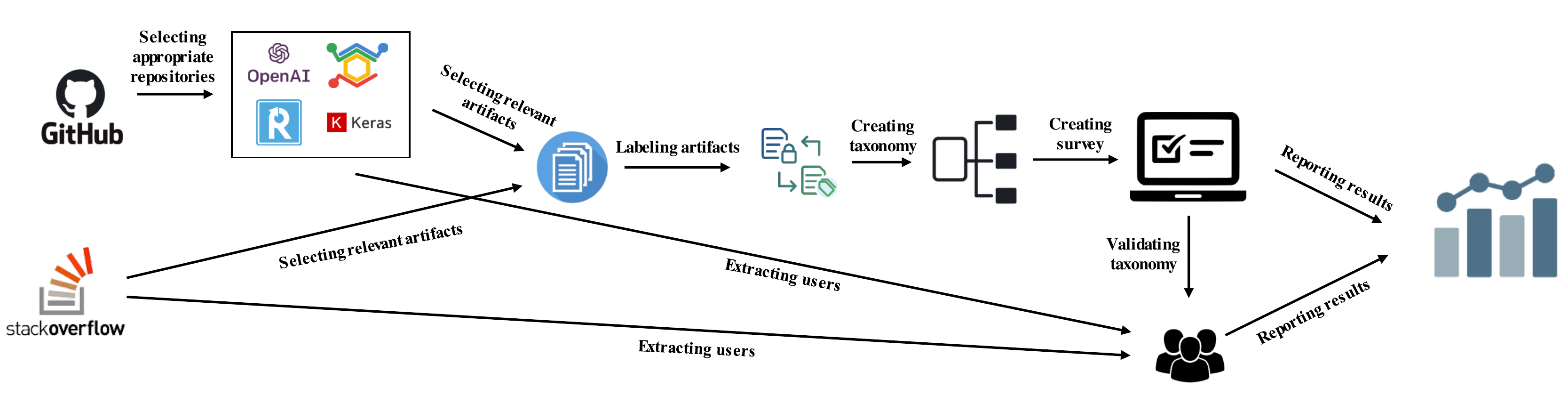}
    \caption{Main steps of the adopted methodology.}
    \label{fig:methodology}
\end{figure*}

In this subsection, we describe the adopted methodology in this paper to construct and validate the proposed taxonomy of faults in DRL programs. Figure~\ref{fig:methodology} presents the main steps of our methodology.
\subsubsection{Manual Analysis of DRL programs}
We considered four popular DRL frameworks: OpenAI Gym, Tensorforce, Dopamine, keras-rl and two main sources of information: GitHub and StackOverFlow. According to GitHub information released in 2019, GitHub has more than 50 million users and about 100 million software repositories \cite{github}. Thus it can be considered as the most important source of open source software artifacts. On the other hand, Stack Overflow is the largest Q\&A website for developers in the stackExchange network. To construct our initial taxonomy, we manually analyzed issues from GitHub and SO discussions related to the four selected frameworks.

\paragraph{Mining Samples from GitHub and SO}
SO posts are our main source of information for bugs/issues related to DRL.
We searched SO questions with five tags: four tags for our targeted frameworks (one tag for each one) and one general tag of \texttt{reinforcement-learning}.
This search returned a total of 2072 posts. Next, we excluded questions without at least one "accepted" answer. This filtering step is important to ensure 
that only questions with a verified answer are analyzed in our study. This process left us with 329 SO posts with at least one accepted answer. We found a total of 1743 SO questions about DRL without an accepted answer. Since this number is large in comparison to questions covering other topics on SO, we manually inspected them to understand the reasons behind the low number of questions with an accepted answer. We made the following observations by analyzing 207 random samples leading to a 95\% confidence level. We categorized such questions into six groups (number in parentheses show relative frequency of each group):
\begin{enumerate}
    \item Basic concepts (20\%),
    \item Without acknowledgment (32\%),
    \item Implementation issues (27\%),
    \item Answered by the owner (2\%),
    \item Irrelevant questions (7\%),
    \item Others (12\%).
\end{enumerate}
Some questions were related to basic concepts of DRL (e.g., SO\#60930232) so they did not receive any answer. Without acknowledgment questions (e.g., SO\#29574444) are those that mentioned relevant issues and received some responses. Although these answers were not accepted, we identified at least one of them as correct that pointed out promising hints but remained without acknowledgment. It seems that users could not recognize the correct answer to assign the accepted answer flag to it. Implementation issues category represents questions that mentioned the possibility of providing DRL using DL frameworks from scratch or applying DRL to some specific problems such as seeking a target by a vehicle (e.g., SO\#4256948 and SO\#56986663). For such questions, there is not an absolute response, so the questions remained unanswered. In some cases, the user who had asked the question answered her own question, but did not assign the accepted answer flag accordingly (e.g., SO\#45364837). We also observed irrelevant questions that have received negative scores from experts and remained unanswered (e.g., SO\#50544568 and SO\#56954306). Finally, if the question could not fall in any of the mentioned categories, we classified them as “Others” (e.g., SO\#63419780).\\
On the other hand, for questions with at least one accepted answer, we observed that the median time to receive a correct answer is about 13 hours (the average is 2.07 days). This time is quite long compared to DL questions that are reported to receive a correct answer in less than 5 hours (on average for difficult questions about performance) \cite{DL_challenges}. It is acknowledged in the literature that the more time it takes for a post to receive an accepted answer, the more difficult the post is \cite{bagherzadeh2019going}. We believe the main reasons for the high number of questions without answers and long time to answer for DRL questions are 1) complexity of DRL-based systems in comparison to DL-based systems and other ML branches and 2) lack of knowledgeable experts to answer questions since DRL is not as popular as DL.\\
We have used GitHub issue tracker to investigate issues of the four DRL frameworks targeted in this study. Usually, bugs and faults related to the development of a framework are mentioned over the repository's issue tracker. But in some cases, developers post bugs found while using a framework over the GitHub issue tracker of the framework. This is why we also referred to GitHub repositories. For each framework, issues likely related to fixing problems were collected. Finally, we identified 761 issues from GitHub issue tracker of four repositories: 151 for OpenAI-gym, 300 for Tensorforce, 200 for Keras-rl and 110 for dopamine. We extracted all issues but we used only the ones labeled as “closed” in our study. This decision aimed to ensure that we analyze only issues that were solved. So, we were left with 432 issues after this filtering.\\
Notably, DRL is an application of DL and consequently the number of reported issues about DRL over GitHub and SO is still less than what one may observe for DL. However, in this study, we have inspected 329 SO posts and 432 issues from GitHub: 761 artifacts have been analyzed totally. Compared to similar studies our dataset is still considerable: a recent taxonomy of faults in DNNs studied programs developed with Tensorflow, PyTorch and Keras by analyzing 1356 artifacts and ending up with 375 relevant artifacts \cite{DLtaxo2020} for all types of bug in DNNs. Another study on bug fix patterns in DNNs inspected 667 samples \cite{islam2020repairing}.

\paragraph{Manual labeling}
We manually analyzed all collected data from SO and GitHub. Similar to Humbatova et. al \cite{DLtaxo2020}, we have used an open coding procedure \cite{seaman1999qualitative}. A shared document including the link to all artifacts have been used to make it possible for all authors to work together during the labeling process. Each artifact was inspected by reading specific parts of its document (code snippet, comment, description) and all related discussion provided by the owner or other users. We removed the underlying artifact from our analysis if: 1) it was not related to a bug-fixing activity, 2) it was related to an issue in the framework itself, not the DRL program, 3) it was related to a common programming error not a DRL-specific bug, and 4) the root cause of the fault was not clear for the authors, based on the associated information of artifact or authors’ analysis. We performed the labeling process in three rounds. For the first round, two authors labelled all the documents by their own defined descriptive label. In this round, we considered only leaf groups by defining a descriptive label without proposing a hierarchical taxonomy. Then, other authors commented on the labelling by proposing a new label where applicable. In this second round, a hierarchical taxonomy was developed by grouping similar leaves. We kept all the labels on our shared document for further discussion and reuse. For the final round, all the labelled documents and the taxonomy were investigated in a team meeting. Conflict resolution was performed in this round. Overall, we found 47 bugs/issues reported by users working with DRL while about 86\% of them were found from SO posts. We ended up with 27 real labelled faults in DRL programs in 11 distinctive categories.

\subsubsection{Building and validating the taxonomy}
We have used a bottom-up approach \cite{vijayaraghavan2003bug} to categorize the faults and construct the taxonomy. First, similar labels are grouped into categories. For each category, we did a double-check by investigating all artefacts in the category to make sure that they are classified correctly similar to labelling rounds. Afterward, parent categories are built based on the “is-a” relationship between each category and its subcategories. During the process, we have discussed each modified version of the taxonomy in virtual meetings with all authors. At last, the taxonomy was finalized in a meeting by exploring all categories, subcategories and leaf nodes.\\ 
A validation process is required to ensure that the taxonomy is well-organized and covers real faults in the DRL program. We conducted a survey to validate the taxonomy involving DRL practitioners/researchers. We targeted practitioners or researchers with a good experience in RL and DRL. Similar to our method for building the taxonomy, we used GitHub and Stack OverFlow to extract information about suitable survey participants. To find participants with a good understanding of RL, we first identified the most popular repositories that focused on RL. Next, we filtered out the ones that do not have software artifacts. After this filtering, we extracted and sorted repositories' contributors based on their activity in the selected repositories. Next, we selected contributors who have been active in updating selected GitHub repositories during the last year. Finally, this process left us with 131 repositories. To find participants from SO, we used the posts extracted during the mining phase, i.e., the posts that have \texttt{reinforcement-learning} tag and at least one accepted answer. Next, we identify the users who posted accepted answers on the selected posts and sort them based on their up-vote and down-vote. Since we could not access the email address of SO users, we carried out a search for each identified user in the web to find their profile and emails from other sources such as GitHub and Google Scholar. In the end, we obtained a list of 210 users; 170 from GitHub and 40 from SO. We sent out the survey by email to all of them and 19 persons participated in our survey which corresponds to a participation rate of 9\%. We received responses from 8 researchers and 11 developers. The minimum coding experience for ML/DL and DRL were “1-3 years” and “less than 1 year”, respectively. The most experienced participant had more than 5 years of experience in both ML/DL and DRL fields while the median for ML/DL and DRL was “3-5 years” and “1-3 years”, respectively.\\
Our survey was created using Google Forms \cite{googleForm}, a well-known online tool for creating and sharing online surveys and quizzes. The survey started by some background questions about job title, DL/DRL-specific programming experience and familiar languages/frameworks. Then, particular questions about our taxonomy were asked. We put distinctive multiple-choice questions for each of our 11 leaf nodes including a description of the corresponding type of fault. In each question, the participants are asked to answer whether they have seen such fault in their own experience or not. The positive answer option asked them to also select a severity level for the fault (minor/major) and the required amount of effort to detect/fix the fault (easy/hard). This allowed us to assess the observed occurrence and the perceived severity of faults provided in our taxonomy, by developers/researchers, at the same time. At the end, we asked if the participant has observed any problems related to DRL that have not been considered in the survey and her availability for an interview to discuss the subject in detail. By these final questions we aimed to find out missing bugs/issues in our presented taxonomy and identify possibilities of further investigation.
\subsection{The Taxonomy}
\begin{figure*}
\includegraphics[width=\textwidth]{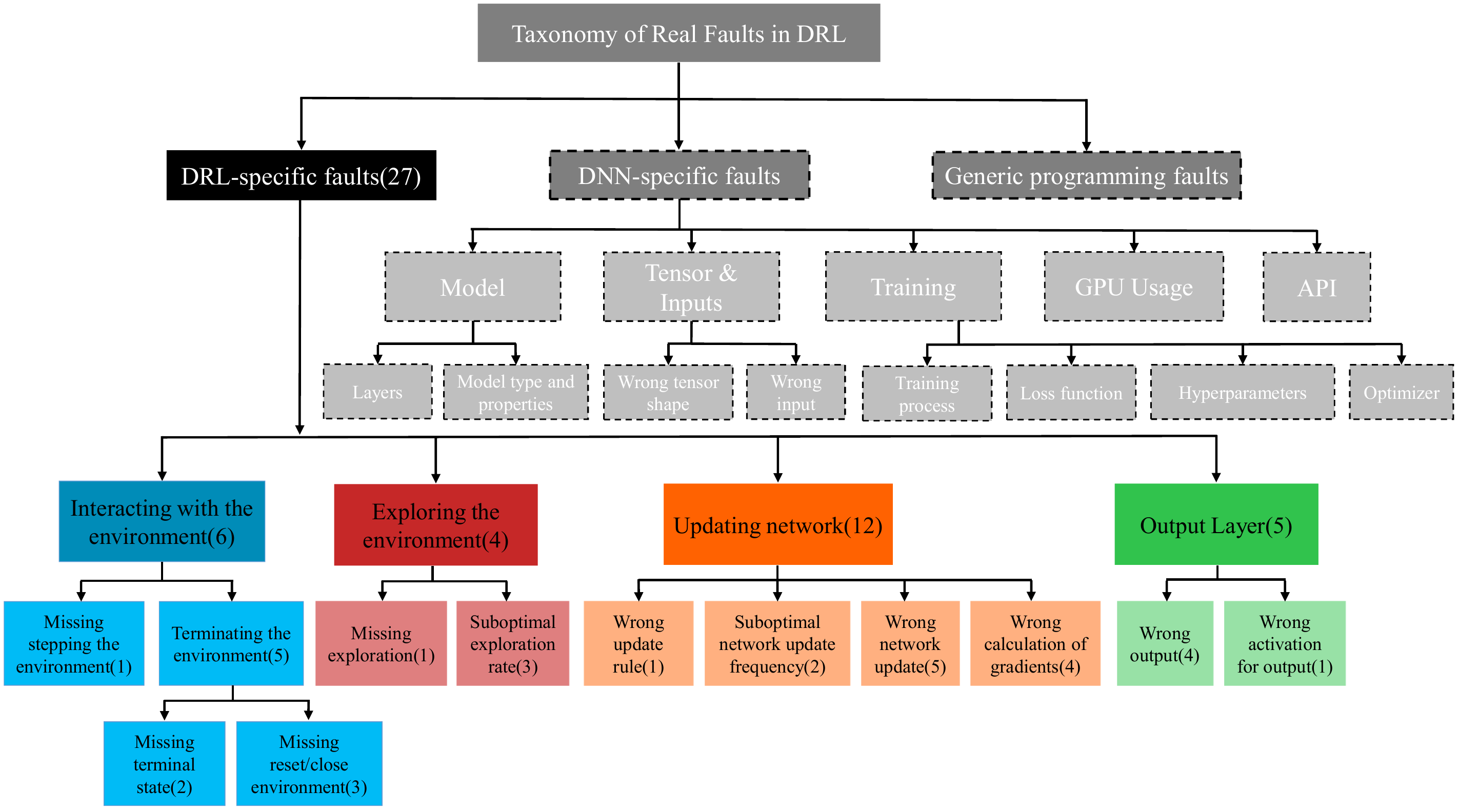}
\caption{Taxonomy of real faults in Deep Reinforcement Learning. Colored boxes are newly added fault types while existing categories are shown in light gray with dotted borders.} 
\label{fig:taxonomy}
\end{figure*}

In general, three types of faults could happen in DRL programs:
\begin{enumerate}
\item Generic programming faults: Generic programming faults: these faults are naturally excluded in our study and similar studies on DL faults \cite{DLtaxo2020,DL_bugs_1,DL_bugs_2},
\item DNN-specific faults: faults related to constructing and training a DNN \cite{DLtaxo2020,DL_bugs_1,DL_bugs_2},
\item DRL-specific faults: these faults occur during the development of a DRL algorithm, i.e, using DNNs in a RL algorithm..
\end{enumerate}
In our study, we did not investigate generic programming faults as explained in Subsection 3.1.1. Since DNN faults have already been studied in some previous works \cite{DLtaxo2020,DL_bugs_1,DL_bugs_2} and may happen in any application of DNNs, we did not investigate them as well. Thanks to a recent interesting research that categorizes observed faults in three popular DL frameworks (Tensorflow, PyTorch and Keras) \cite{DLtaxo2020}, we can discuss these faults systematically. This recent taxonomy has five categories: Model (type of layers and architecture of DNNs), Tensors and Inputs (size and type of input data), GPU Usage, Training (training process of DNNs like loss and optimizer) and API (usage of framework APIs). In fact, the taxonomized faults could happen when constructing, training and using a DNN regardless of the application domain. For example, selecting a wrong architecture for a network, missing a specific layer, wrong initialization of network’s parameters or wrong loss definition. Instead, in our proposed taxonomy, we have categorized DRL-specific faults which are related to using DRL algorithms, i.e., employing DNNs in RL, and can only happen in DRL programs. The focus of this study is then on DRL-specific faults since to the best of our knowledge, “faults related to developing a DRL algorithm” have not been reported in the literature previously. Moreover, according to the manual analysis of buggy programs and the feedback we received in our survey for validating the taxonomy, we believe that DRL-specific faults are important to study. The application of DRL in various domains is growing, so developers of DRL programs should be aware of various types of fault occurring in these programs.\\
However, we have blended our proposed taxonomy of DRL faults with the taxonomy of DNN faults \cite{DLtaxo2020} to 1) make the taxonomy self-contained in terms of faults that may happen in DRL programs and 2) help the reader to understand the significance and hierarchy of various faults. It should be noted that there is an exception for faults related to training data. Since in RL, an agent is interacting with an environment and the learning process is based on information gathered through interactions, traditional “training data” which has been used in classification tasks is not relevant to DRL programs. Hence, we removed categories related to “training data” including preprocessing and quality issues. Figure \ref{fig:taxonomy} illustrates the final taxonomy. Colored boxes are newly added fault types while existing categories are shown in light gray. The DRL-specific sub-taxonomy includes 4 categories and 11 leaves. The numbers after each title represent the number of posts assigned to that title during manual labelling. In the following, we first briefly describe DL faults reported in \cite{DLtaxo2020}. Then, our proposed categories of DRL faults will be discussed.
\subsubsection{DL faults}
According to \cite{DLtaxo2020}, there are five main categories of faults in DL programs: Model, Tensors and Inputs, GPU Usage, Training and API. Any fault related to the structure and properties of the DL model is classified as \textbf{Model}. It has two subcategories of \textit{Model type and properties} and \textit{Layers}. Faults related to the model as a whole not its components are categorized as \textit{Model type and properties}. For example, wrongly selected model for a task, too few or too many layers in a model which is identified as suboptimal network structure, or wrong initialization of weights in the network. \textit{Layers} covers faults related to a particular layer as a component of a model, e.g., missing average pooling layer after convolution layer(s), missing activation function or mismatch of layers’ dimensions.\\
\textbf{Tensors and Inputs} gathers issues related to a wrong dimension, type or format of data processed by the network and has two subcategories: \textit{Wrong tensor shape} and \textit{Wrong input}. The former covers bugs raised by operation on tensors with incompatible or incorrectly defined shape, e.g., using a transposed tensor instead of the normal one. Incompatible format, type or dimension of input data to a layer or methods result in faults of \textit{Wrong input}.\\
All faults that are encountered during the training process of DL programs are categorized as \textbf{Training}. \textit{Hyperparameters} covers problems due to tuning the hyperparameters of the DL model including learning rate, number of epochs and batch size. Wrong calculation, wrong selection or missing a loss function leads to faults from the subcategory of \textit{Loss function}, which affect the effectiveness of learning algorithms. \textit{Optimizer} includes faults like wrong selection or bad parameter setting of the model training optimizer. Faults such as attempting to fit a too big model into memory or to refer to non-existing checkpoint during model restoration fall in the subcategory \textit{Training model}.\\ 
Usually, DL frameworks employ GPU devices to run the code, so the \textbf{GPU Usage} subcategory includes faults that occur while using such devices. For instance, wrong reference to a GPU device or faulty data transfer to a GPU device. Finally, failures related to 
the wrong usage of framework APIs (i.e., using an API in a way that does not comply with its definition) fall into the 
\textbf{API} subcategory. It should be noted that none of the extracted faults in this paper could fall into categories of the mentioned taxonomy of DL faults. 
\subsubsection{Interacting with the environment}
In a DRL program, an agent must interact with an environment to learn. Usually in each DRL program, there are some parts for initializing, retrieving current state, submitting action and receiving next state and the reward. DRL frameworks support some well-known environments with the possibility of defining new environments. They provide developers with APIs to work with the environments. This category considers faults that occur when a DRL program tries to interact with the environment.

\paragraph{Type 1: Missing stepping the environment}
Failure to timely push the environment to a new state and get the associated reward lead to problems in learning. During each episode of interaction between an agent and its environment, the environment must be moved to a new state getting the reward associated with the transition. Otherwise, the agent would lose the track of reward or sequence of the environments' states.\\ \\
\textbf{Terminating the environment:} This subcategory includes faults related to the ending of an episode of interaction with the environment or restarting the environment for the next episode. 
\paragraph{Type 2: Missing terminal state} This category contains problems related to missing or wrongly detected terminal state of the environment. Finally, each episode of agent interaction with its environment should be terminated normally by reaching the terminal state. Missing this state may lead to inefficient learning. For example, using default definitions of the framework's API for detecting the terminal state leads to such fault. 
\paragraph{Type 3: Missing reset/close environment}
It includes problems related to missing or bad termination (and restarting) of each round of agent interaction with its environment. At the end of each episode, the environment must be properly closed or reset to its default configurations for the next episodes. Actually, execution of the next episode and correctness of state sequences depend on the successful termination of the previous episode. Wrongly positioned API call for resetting environment is an example of this type of faults.

\subsubsection{Exploring the environment}
A RL agent must prefer high-reward actions that have been tried previously to obtain more reward. On the other hand, to detect such actions, the agent must investigate actions that have not been taken before. The agent not only has to exploit its experience of actions with higher reward, but also has to explore to improve action selection in the future. The dilemma is that neither exploration nor exploitation can be successful exclusively without failure. A variety of actions has to be attempted by the agent, gradually favoring those that appear to be best. Although there are various methods to perform exploration in DRL, enough exploration of the environment is crucial for an effective performance. Faults in this category fall into two groups: missing the exploration phase in the code and choosing suboptimal exploration rate. 
\paragraph{Type 4: Missing exploration}
This tag is identified as the failure to explore the environment in the case that it is necessary according to the algorithm. Lack of exploration leads to poor performance of the algorithm in terms of mean reward. Sometimes developers rely on the output of the neural network (e.g., DQN) to have enough flexibility to cause sufficient exploration but using explicit methods like epsilon greedy is more effective. 
\paragraph{Type 5: Suboptimal exploration rate}
It is widely acknowledged in the RL community that balancing exploration and exploitation is crucial to achieving good performance \cite{sutton2018reinforcement}. This tag covers problems related to suboptimal exploration parameters (e.g., epsilon in epsilon-greedy method) or suboptimal decay rate that leads to poor performance of the algorithm.

\subsubsection{Updating networks}
This is the largest category in our taxonomy. It includes faults related to updating DNNs in DRL programs. According to the adopted DRL architecture/algorithm, various updating or training procedures should take place in DRL programs, e.g., training Q-network and updating target network in DQN, policy and value networks in policy gradient algorithms, and updating replay buffer.
\paragraph{Type 6: Wrong update rule}
A new experience from the environment should be incorporated into the existing experiences by an update rule. This leaf covers issues related to using an incorrect update rule for value or policy function including suboptimal learning rate and wrong implementation of the update rule.
\paragraph{Type 7: Suboptimal network update frequency}
Problems related to suboptimal update frequency of networks’ parameters (including the target network) leading to unstable learning and increasing loss value are categorized in this group. For example, in DQN, the target network must be updated frequently, so choosing a too high update rate can lead to unstable learning process.
\paragraph{Type 8: Wrong network update}
This tag covers faults related to the wrong update of networks or its parameters. Each network or set of parameters (like Q and target network in DQN or policy and value networks in policy gradient algorithms) must be properly updated based on new values or recent observations. Examples are update the wrong network, wrong update statement and missing the update statement(s).
\paragraph{Type 9: Wrong calculation of gradients}
Problems related to wrong calculation of gradients for learning, including computation of one network's gradients with respect to another network's are categorized in this tag. Since in some complicated DRL methods, different networks must be trained according to the output of other networks or returning reward from the environment, it is a quite frequent fault in such methods.

\subsubsection{Output layer}
When DL is applied to typical ML problems like classification, the output of the network is much more familiar in comparison to RL problems. A particular action, a vector of state-action values or a probability distribution over possible actions could be the output of DNNs in DRL programs. Due to this diversity, faults related to the output layer of the network are observed in our study. We have categorized these faults in two leaves.
\paragraph{Type 10: Wrong output}
This tag includes issues related to failure to define a correct output layer for the network with respect to the environment and algorithm. This type of faults leads to issues in determination of actions and poor learning.
\paragraph{Type 11: Wrong activation for output}
Failure to define a correct activation function for the output layer leading to incorrect action determination is labelled in this group. As a real example, if sigmoid activation function is used when the output is expected to be a reward value, the activated output range will be different from the original reward.

\subsection{Validation results}
Table \ref{validationTable} shows the results of the validation survey for the proposed taxonomy. The ratios of answers for each category are reported. The answers reveal whether or not the participants ever faced the related faults, their perceived severity for each type of fault (as “minor/major”), and their estimation of the effort required to fix the faults (as “easy/hard to fix”). The results confirm the relevance of all categories in the taxonomy since all of them have been encountered by at least 37\% of participants. \textit{Exploring the environment} is the most popular category; it has been experienced by 89\% of participants. 
Participants have determined \textit{Missing exploration (Type 4)} as the most critical faults by 63\% of answers (highest rate for "Yes, major and hard"). The least popular type is \textit{Missing stepping the environment (Type 1)} experienced by 37\% of all the survey participants which is a non negligible fraction. The average of answers that included “yes” is 71\%. This result shows that the taxonomy covers relevant types of faults that have been experienced by most of the participants as DRL developers/researchers. However, two types of faults received more than 50\% “no” answers as the least observed type of faults in DRL. Consequently, participants validated 9 categories out of 11 on average, the ratio of categories that received more than 50\% of responses that included “yes”. 

In response to our question about the completeness of our taxonomy, some participants mentioned examples of faults they thought we missed in the reported taxonomy:
\begin{enumerate}
\item Generic coding problems were commented by two participants (an example is: \textit{"Wrong data type when storing the environment state somewhere"}),
\item The effect of the fault was described in two others rather than the root cause of symptom (for example, \textit{"diverging behaviour and instability"} and \textit{"not converging entropy/gradients"})
\item One reported faults related to DL which are not specific to DRL (such as \textit{"neural network initialization"}). 
\item Three participants commented on the definition and formulation of a DRL problem including evaluation metrics and reward function which are not directly related to faults in DRL programs (e.g. \textit{"Problems related to defining reward systems for new environments for efficient learning."}). 
\end{enumerate}
\begin{table*}
\caption{Results of validating survey for the proposed taxonomy.}
\label{validationTable}
\begin{center}
\resizebox{\textwidth}{!}{
\begin{tabular}{|c|c|c|c|c|c|}
\hline

\multirow{3}{*}{\textbf{Faults Type}} & \multicolumn{5}{c|}{\textbf{Responses}}\\
& No & Yes, & Yes, & Yes, & Yes, \\
&  &minor and easy&minor but hard&major but easy&major and hard\\

\hline

\hline
Type 1 & 63\% & 16\% & 5\% & 11\% & 5\% \\

\hline
Type 2 & 42\% & 16\% & 16\% & 21\% & 5\% \\

\hline
Type 3 & 37\% & 53\% & 0\% & 11\% & 0\% \\

\hline
Type 4 & 11\% & 5\% & 5\% & 16\% & 63\% \\

\hline
Type 5 & 11\% & 16\% & 16\% & 16\% & 42\% \\

\hline
Type 6 & 11\% & 21\% & 11\% & 26\% & 32\% \\

\hline
Type 7 & 16\% & 0\% & 26\% & 21\% & 37\% \\

\hline
Type 8 & 16\% & 0\% & 26\% & 21\% & 37\% \\

\hline
Type 9 & 32\% & 32\% & 5\% & 16\% & 16\% \\

\hline
Type 10 & 53\% & 32\% & 0\% & 11\% & 5\% \\

\hline
Type 11 & 42\% & 32\% & 5\% & 21\% & 0\% \\
 
\hline
\end{tabular}}
\end{center}
\label{table:table-results-1}
\end{table*}
\subsection{Discussion}
\label{discussion}
The number of research works in software engineering that use SO as a source of information is growing which raises some concerns about quality \cite{meldrum2017crowdsourced}. The problem is the validity of its utility and reliability which has not been investigated yet. Humbatova et al. \cite{DLtaxo2020} also reported the views of some of their SO interviewees stating that Q\&A over SO  does not reflect the problems developers/researchers faced when developing DL programs. To compensate for this issue, we have excluded non-relevant artifacts during our manual analysis, have investigated each artifact by participation of more than one evaluator and have conducted a survey to validate the results.
Participants in our online survey have provided us with some comments. We present and discuss common problems among these comments:
\begin{enumerate}
  \item Reproducibility of result : Stochastic nature of RL problems make it difficult to reproduce the results and investigate the correctness of implementation. One of them mentioned that: \textit{“finding a good random seed is annoying, even worse is the high sensitivity to different seeds: performance may greatly vary from one seed to another”}. Another participant stated that sometimes one would not get good results just because of bad luck in a run.
  \item Complexity of DRL frameworks: some participants complained that DRL frameworks are quite complicated due to the modular design. So, they prefer to 
  to write all the code from scratch, one participants noted that \textit{"Writing all code from scratch seems to be the only way to have full control of what's going on. I believe we can do better"}. This makes it difficult to develop, debug and test DRL algorithms. They stated the  advantages of single-file implementations to be their easy development and debugging. Moreover, they expected to see new frameworks including such features (e.g., \textit{"I am a huge advocate for single-file implementations. Easier to debug, inspect, and develop new algorithms."}). 
  \item Scalability challenge: based on participant views, it seems that scalable experiment is a challenge in DRL programs. More effort is required to address this issue, for example to handle a reasonable number of agents in multi-agent problems.
\end{enumerate}

\section{Model-based Fault Detection in DRL Programs}
\label{modeling}
In this section, on the top of the proposed taxonomy, we aim to automatically detect bugs in models of DRL programs by model-based static analysis. To do so, a meta-model for DRL programs is proposed. In this first attempt, we consider DQN which is a well-known DRL algorithm. This meta-model captures the base skeleton and fundamental properties of DRL programs independent of available libraries. Then, a detection mechanism is proposed using graph transformations.

\subsection{A Meta-model for DRL Programs}
\begin{figure*}
\begin{center}
\includegraphics[width=0.95\linewidth]{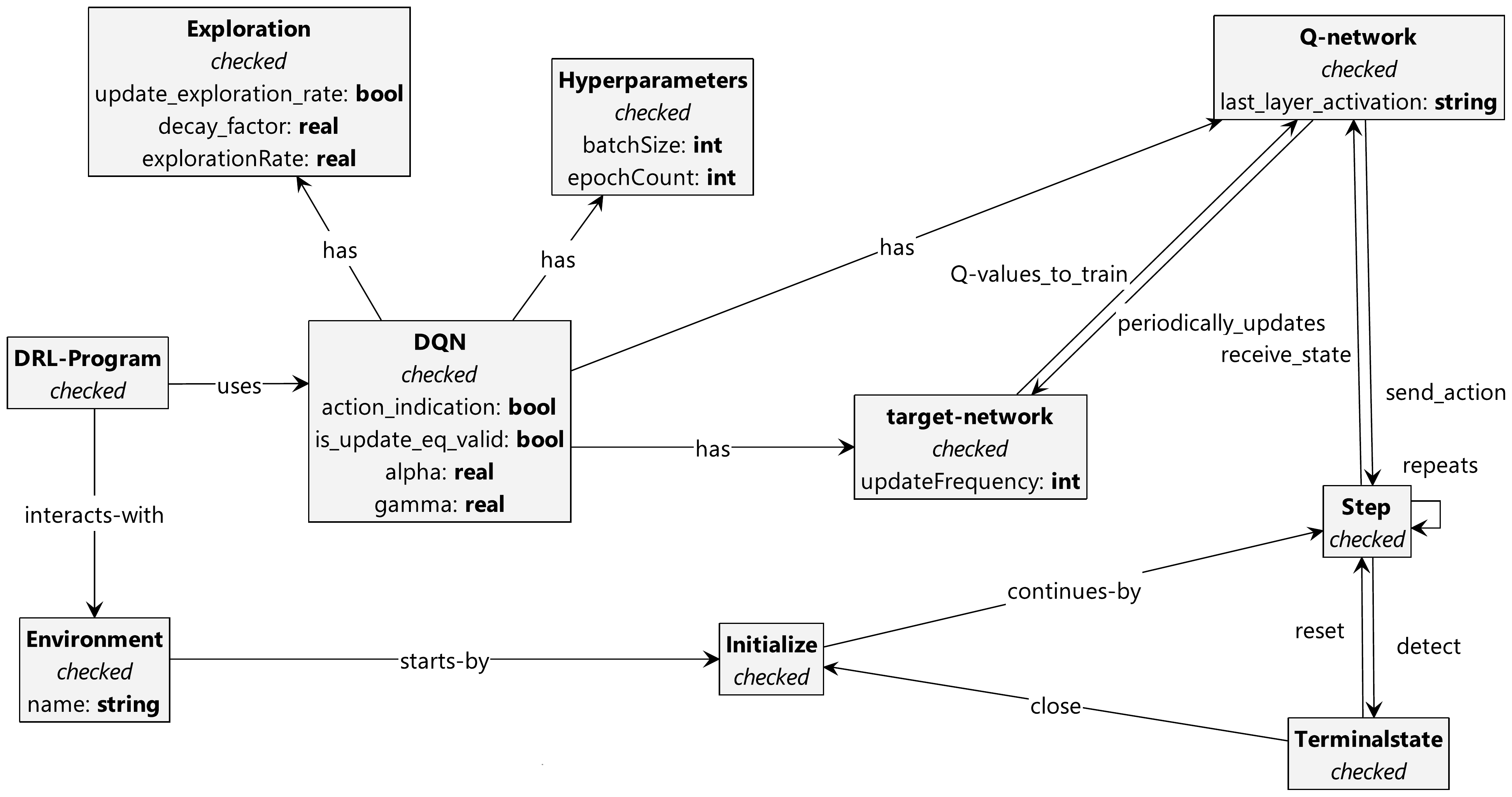}
\caption{The proposed meta-model (type graph) for DQN-based DRL programs.}
\label{fig:model}
\end{center}
\end{figure*}

In this subsection, we want to address the following questions: is there any generic representation of DRL programs that is independent from employed development libraries? or, can we define a meta-model of DRL programs and how can we model a DRL program? In \cite{hartmann2019meta}, researchers proposed a meta-model for meta-learning. They presented an overview of the meta-learning concepts –on a meta-modelling level– with possible variabilities and discussed how their meta-model could be integrated into existing modelling frameworks and tools. However, while their meta-model includes ``Learning Block", ``Learning Algorithm", and ``Hyperparameters", no further details were presented and they did not explore the possibility of identifying faults in ML models. In this section, we present a particular meta-model for DRL programs and an approach for modeling such programs. Afterward, one could perform static analysis of models of DRL program for detecting faults in them. We believe that a generic meta-model for DRL programs would be helpful for understanding DRL programs written by developers using third-party libraries.\\
A DRL program is basically an RL program that adopts DNNs. A typical RL program has two main components: a component to interact with the environment and a component to learn the optimal policy and make decisions upon the environment. In a DRL program, one or multiple DNNs are used for learning policy and selecting actions. Normally, a feedforward multilayer perceptron (MLP) architecture is used for DNN in DRL programs. Like other computational models, DNN attempts to find a mathematical mapping from the input into the output during a learning phase. Usually, a set of inputs and desired outputs (or targets) is provided for learning which is called Dataset. In the case of DRL, input is agent's observations (including state of the environment, reward and agent's actions) and the output could be agent's action, estimated reward or its policy. We have briefly reviewed various DRL approaches in Section \ref{drl}. For the sake of simplicity, in this section we consider the DQN algorithm for our meta-modeling approach, since it is a well-known approach in DRL. The DQN is a value-based DRL that employs a DNN, namely Q—network, to perform a nonlinear approximation which maps the environment state into an action value.\\
Since we have used a graph-based approach for modelling and detecting faults in DRL programs, here we briefly review its important concepts. Graph transformation system (GTS) \cite{heckel2006graph} (also called graph grammar) is a formal language for the specification of software systems, in particular those with dynamic structures. The definition of an attributed GTS consists of a triplet \textit{(TG, HG, R)} in which \textit{TG} is a type graph, \textit{HG} is a host graph, and \textit{R} is a set of rules for graph transformation. \textit{TG} is defined by four components, \textit{$TG=(TG_N, TG_E, src, trg)$}. \textit{$TG_N$} and \textit{$TG_E$} includes all node types and edge types respectively. \textit{$src: TG_E \rightarrow TG_N$} and \textit{$trg: TG_E \rightarrow TG_N$} are two functions that determine the source/destination nodes of an edge, respectively. The initial configuration of a system specified by GTS is presented by the host graph which is an instance of the type graph. Therefore, each component of the host graph, node or edge, must have a component type in the type graph. A host graph \textit{HG} may instantiate from a type graph \textit{TG} using a graph morphism function \textit{$type_G: HG \rightarrow TG$}, in which the components of \textit{HG} are instantiated from \textit{TG}. Other configurations or states of a system are generated by successive applications of transformation rules on the host graph. A transformation rule \textit{r} in \textit{R} is defined by a triplet \textit{$(LHS_r, RHS_r, NAC_r)$} in which \textit{$LHS_r$} (left-hand side) represents the preconditions of the rule whereas \textit{$RHS_r$} (right-hand side) describes the postconditions. Moreover, there may be a Negative Application Condition (NAC) for the rule \textit{r}, meaning that the rule \textit{r} can be applied only when \textit{$NAC_r$} does not exist in the host graph. By applying the rule \textit{r} to the host graph \textit{HG}, which is an instance model of the meta-model or type graph, a matching of the \textit{$LHS_r$} in \textit{HG} is replaced by \textit{$RHS_r$}. Formally, a graph morphism exists between \textit{$LHS_r$} and the instance model \textit{HG}. The application of a rule is performed in four steps: (1) find a matching of \textit{$LHS_r$} in \textit{HG}, (2) check \textit{$NAC_r$} that forbid the presence of certain nodes and edges, (3) remove a part of the host graph that can be mapped to \textit{$LHS_r$} but not to \textit{$RHS_r$}, and (4) add new nodes and edges that can be mapped to the \textit{$RHS_r$} but not to the \textit{$LHS_r$}.\\
Our meta-model for DQN-based DRL programs includes two main parts: Environment and DQN as decision making component. The proposed meta-model is represented by a type graph. This type graph is illustrated in Figure \ref{fig:model}. The node representing the \textbf{DRL-program} has two edges to \textbf{Environment} and \textbf{DQN} indicating its main components. It should be noted that our aim of meta-modeling is the detection of faults in DRL programs; therefore the most relevant components have been incorporated into the meta-model. Since we focus on DRL-specific bugs in this paper, structure and parameters of DNNs are not presented in our meta-model explicitly.
\subsubsection{Environment}
A DRL program interacts with an environment. \textbf{Environment} contains variables for number of actions and number of states. The program should start with initializing the environment. Then, it continues to \textbf{Step} the environment. After initialization, the agent starts interacting with the environment by perceiving the current state of the environment, making decisions to perform an action and executing the action in the environment. Afterward, the agent might receive a reward and the environment updates its state. We call this ``stepping the environment" which could be repeated during an episode. Finally, an episode of agent-environment interactions should terminate by detecting a terminal state. The agent may reset the environment for starting another episode or close it at the end.

\subsubsection{DQN}
A DQN-based DRL program normally employs a DNN which is called \textbf{Q-network} to approximate a vector of state-action values by receiving environment's state as input. It may benefit from another DNN, namely \textbf{Target-network}, to instantiate fitted Q-network for some iterations and applying network’s parameters update only periodically not in every iteration. \textbf{Q-network} receives the current state and reward of the environment from \textbf{Step}, calculates the best possible action and sends it back to \textbf{Step}. Meanwhile, \textbf{Target-network} provides \textbf{Q-network} with Q-values for training and \textbf{Q-network} updates parameters of \textbf{Target-network} periodically. On the other hand, \textbf{Exploration} is crucial for a successful DQN algorithm. For example, epsilon-greedy simply balances exploration (selecting a random action) and exploitation (selecting actions according to Q-networks) by choosing between these two randomly. In this algorithm, where epsilon refers to the probability of choosing to explore, exploitation occurs most of the time with a small chance of exploring. As time goes by, epsilon decreases by a decay factor. \textbf{Hyperparameters} include parameters like batch size, epoch count, and size of replay buffer.  

\subsection{Modeling DRL programs}
A model of a DRL program includes components that form its source code. There are two ways to build a model of DRL programs: configure an arbitrary model directly or transform a DRL program to a model. Practically, a model could be configured according to a DRL program that has already been developed by a programmer. To do so, the source code of a DRL program should be parsed and converted to a model, which is an attributed graph. Due to the complexity of developing DRL programs and various components which must be identified to build the model, analyzing DRL codes to extract the model is not straight-forward and easy. Hence, as a preliminary approach, we have developed a dedicated converter to transform a sample DRL program to its model in this paper.\\
The meta-model is generic enough to be independent of any specific DRL library or framework. Hence, once we can have a model of a DRL program that conforms to the meta-model, further investigations on the model become possible, such as bug detection. Apart from the work and analysis that are presented in the rest of this paper, we believe that this meta-model can be very useful to understand DRL programs written by third parties. It will be also helpful in understanding the development activities of DRL practitioners; the way they develop DRL programs and the type of faults that they experience.

\subsection{Detecting Faults by Graph Transformations}
\begin{figure*}
\centering
\includegraphics[width=0.8\textwidth]{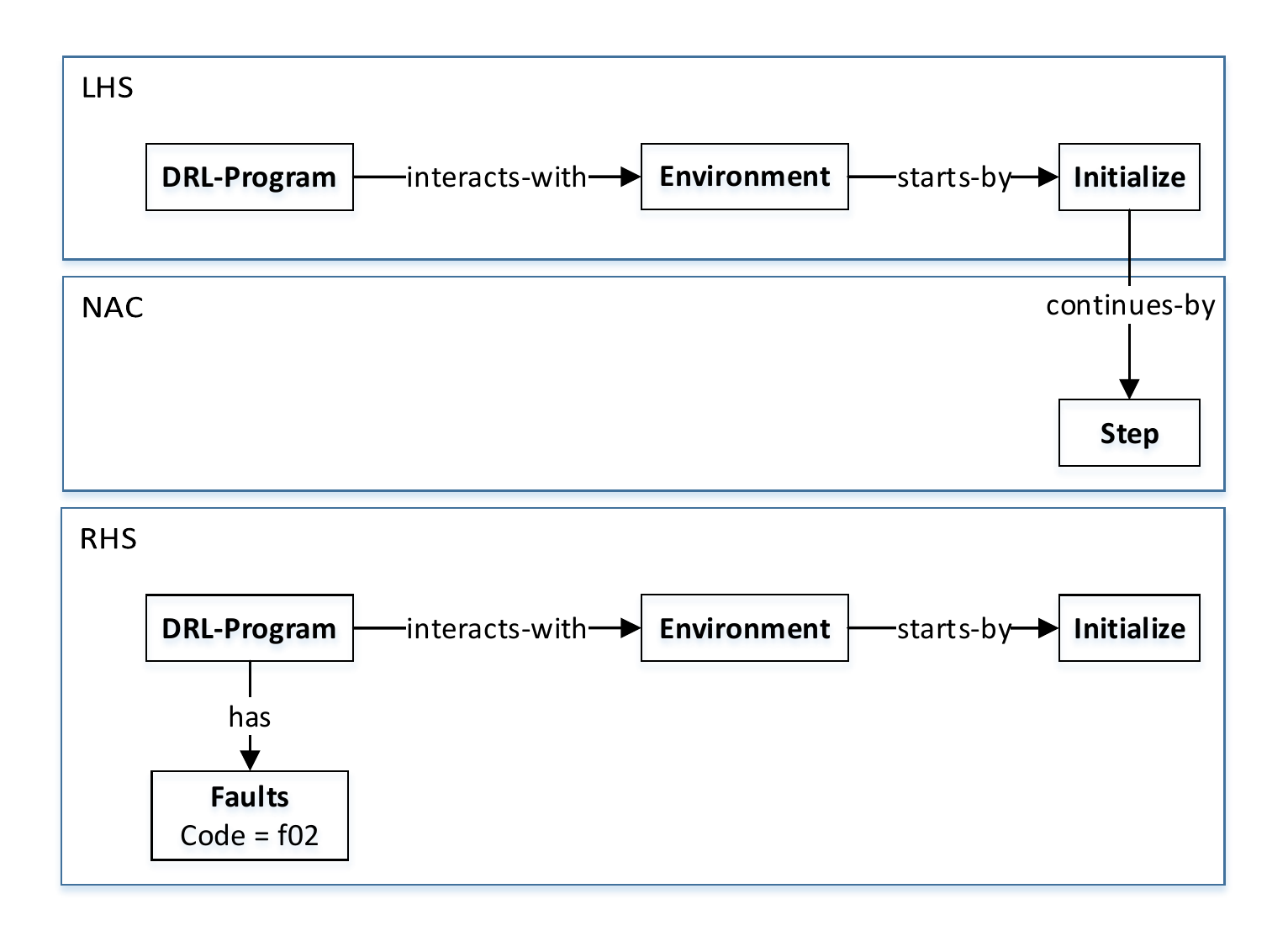}
\caption{An Example of Graph Transformation Rules: Detecting  ``Missing stepping the environment".}
\label{fig:rule}
\end{figure*}

In this paper, the meta-model is presented as a type graph and each model, instantiating the type graph, is a graph extracted from a DRL program. As a straightforward approach, graph transformations are chosen to detect faults in models of DRL programs. For each type of fault, one graph transformation rule is implemented to detect the fault. In fact, graph transformations are used to detect possible faults in a model acting as graph checking operators. A transformation is applicable where the corresponding fault occurred in the model. A fault in the graph (model) could be detected as a missing node, edge or wrong value of a variable. While detecting a fault in a model, some specific conditions would be checked by finding a match of LHS of the rule in the graph and/or the absence of NAC. Once a graph operation is applied, i.e., detecting a fault in a part of the graph, a specific fault code is added to the node or edge in which the violation occurred. This action is represented by the RHS of the rule. For example, Figure \ref{fig:rule} illustrates a transformation rule for detecting ``Missing stepping the environment" fault showing LHS, RHS, and NAC. This fault occurs when the developer forgets to push the environment to a new state and to get the associated reward. LHS shows \textbf{DRL-program} with its initialized \textbf{Environment}. The fault is detected if there is not a \textbf{Step} node just after \textbf{Initialize}. So, NAC forbids the existence of \textbf{Step} right after \textbf{Initialize}. If the fault is detected, RHS adds a \textbf{Faults} node with relevant fault code to the \textbf{DRL-program}. Because of space limitation, we cannot present in the paper all the graph transformations implemented for fault detection. We refer interested readers to the source code of \tool{} which is available online \cite{drlinter}.

\subsection{Implementation}
\label{imp}
The pseudocode of \tool{} is presented in Algorithm \ref{algo:algo1}. The inputs are DRL program and a set of graph transformations rules. At first, the DRL program is modeled as a graph that conforms to the proposed meta-model, i.e., type graph. Then, a checking process runs to find potential faults in the model. This process attempts to apply rules to the graph and stops when no further rule application is possible. Finally, \tool{} traverses this graph to generate a report for the user, containing a description of the faults found for each component of the program. We discuss each step in detail in the rest of this subsection.\\
In \textit{convertDRLProgram}, the source code of a program is parsed to extract relevant information that is necessary to build the model. 
In the first step, we use the Abstract Syntax Tree (AST) to parse the DRL program script. AST represents the abstract synthetic structure of the scripts as a tree. Due to the fact that AST provides the abstract information regarding the scripts and omits very detailed information of code, we used this tool to generate graphs of DRL codes. To cope with generating readable graphs by graph checker, we extract the required abstract information from the DRL script using AST and after that, we start to add the relevant nodes and edges based on the extracted data by AST. The current version of \tool{} works for synthetic DRL programs developed using OpenAI Gym and TensorFlow libraries. Therefore, the API of these frameworks is the main source to detect key parts of a DRL program. In each DRL framework, there are specific APIs for initializing the environment, receiving the current state of the environment, sending the action and working with DNNs including feeding input, calculating output, and training. In this way, the most important parts in the DRL program like Q-networks, target network, exploration rate, and update equation could be identified by the parser. The current version can be extended to work for DRL programs developed by other DRL frameworks.\\
Once DRL source code is modeled as a graph and we have the transformation rules, the detection rules can be used by a graph transformation tool that executes the sequence of graph transformations over a model of the DRL program by calling \textit{graphChecker}. In this paper, we have used the GROOVE toolset \cite{rensink2004groove} to perform graph operations. GROOVE is a tool for implementing, simulating, and conducting analysis of graph transformation systems. It is capable of exploring recursively and collecting all possible rule applications over a start graph. Furthermore, it has a graphical interface for editing graphs and rules, and for exploring and visualising the graph transformation which could be called via command line, as well. The output of GROOVE is called the final graph on which no further rule application is possible. The detection rules are implemented as graph transformations using GROOVE. Each graph transformation could be applied to a graph if there is a fault in the graph. The rules are implemented in such a way that start from the \textbf{DRL-program} node and proceed to the parts of the model. At first, the general structure and connectivity of nodes is tested to ensure that they are well-formed and connected. These transformation rules mark the graph components (nodes and edges) with relevant flags to indicate the performed tests. Then, each graph operation checks specific conditions that are asserted in its rule using the information provided in the graph. A transformation should be fired if a fault is detected in the model of a DRL program. If there are multiple faults or various instances of a fault in the considered model, all of them will be detected by applying multiple enabled rules.\\
Finally, by calling \textit{extractReport}, a report is extracted from the output of \textit{graphChecker}. A simple parser is developed to process the final graph and extract information about detected issues to generate a report for the user. For more details, we refer interested readers to the source code of \tool{} which is available online \cite{drlinter}.
\begin{center}
\begin{algorithm}[t]
\KwIn{A DRL program, \textit{program}, and \textit{rules} as graph transformations}
\KwOut{List of detected bugs of the program}
\SetAlgoLined
$graph \leftarrow{} \text{convertDRLProgram(\textit{program})}$\\
$graph \leftarrow{} \text{graphChecker($graph$, \textit{rules})}:$\\
\begin{enumerate}
 \item starting by $graph$, apply enables rules.
 \item terminate when there is no applicable rule.
 \item \Return $graph$.
\end{enumerate}
$report \leftarrow{} \text{extractReport($graph$)}$\\
\Return $report$
\label{algo:algo1}
\caption{\tool: Model-based Fault Detection in DRL Programs by Graph Transformations}
\end{algorithm}
\end{center}
\subsection{Application scope}
RL, and consequently DRL, basically aims to handle the automatic learning of optimal decisions over time. Agents aim to learn good behavior by modifying their current behaviour incrementally or attempting new ones. Trial-and-error experiences are employed leading to frequent interactions with the environment and information collection. Hence, RL algorithms usually should be designed and tuned depending on the environment, prior knowledge and particular learning objectives for each domain. Therefore, there is not a universal principle for RL and DRL algorithms that can be applied to all problems. In this paper, we investigated real faults in DRL programs. According to our findings, we then proposed a taxonomy of faults in DRL programs. Finally, we proposed an approach to detect these faults in the DQN algorithm, which is a well-known approach in DRL. Therefore, the applicability of the detection approach is limited to DQN while the taxonomy covers faults that might occur in other DRL approaches, not only DQN.\\
\tool{} detects potential faults in DRL programs which is, to the best of our knowledge, the first attempt in this field. The developers provided with information about the detected faults will be able to manually correct them considering the context and the domain of the problem. For example, when \tool{} reports a fault on exploration rate, it draws the developer's attention to a potential issue in exploring the environment that should be investigated. The developer should do inspection accordingly and revise the mechanism or rate of exploration. To automatically suggest repairs for the identified faults, we anticipate that a dynamic analysis of the DRL program will be required in some cases, i.e., running the program and analyzing the training process during runtime, therefore, as future work, we plan to develop fault correction tools that are based on a dynamic analysis of the DRL program. Such analysis will investigate each fault in its own context.\\
\tool{} is a model-based approach, so at first, a DRL program should be modeled as a graph that conforms to the proposed meta-model. The detection rules are then applied on the extracted model to find potential faults. Our current meta-model is designed for the DQN algorithm, a well-known and widely used deep RL algorithm. \tool{} can be easily extended to other deep RL algorithms by proposing a proper meta-model for each algorithm while the rules remain the same. On the other hand, fault detection happens on the model of the DRL program, not the program itself. Therefore, by extending the converter (\textit{convertDRLProgram} function in Subsection 4.4) to extract models from programs that use DQN algorithm but written by other frameworks, \tool{} will be able to detect faults in them using the same meta-model and the same rules.

\section{Experimental Evaluation}
\label{exp}
To evaluate the effectiveness and accuracy of \tool{}, we use a set of synthetic and real-world faulty DRL programs. We also need some real-world faulty samples to imitate the faults occurring in them for creating realistic synthetic examples. In fact, we need some buggy DRL codes that contain the types of faults we covered in \tool{}. Moreover, we need a full DRL program to construct a complete model of the program on which the fault detection could be performed. In the rest of this section, we describe our methodology to find real faulty DRL samples, preparing synthetic faulty examples and the results of fault detection by \tool{} on these examples.
\subsection{Faulty programs}
\begin{figure*}[t]
\begin{center}
\includegraphics[width=0.8\linewidth]{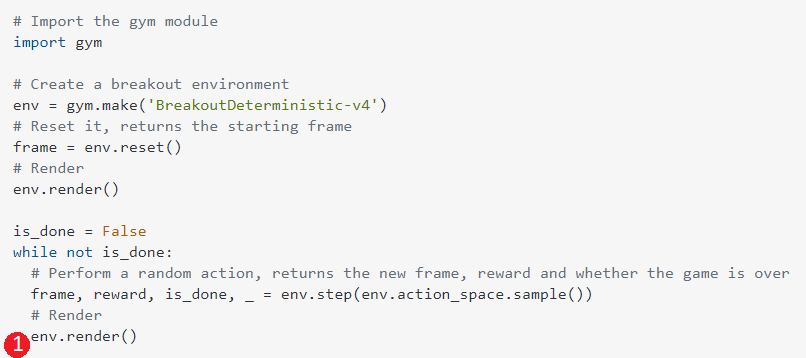}
\caption{Example of a faulty DRL program from SO\#50308750. The developer missed \texttt{env.close()} at the end of the program.}
\label{fig:snippet2}
\end{center}
\end{figure*}

We have used some of the real faulty DRL programs gathered from SO and GitHub that we manually analyzed for constructing the taxonomy to evaluate \tool{}. Anyway, finding faulty DRL programs from the SO posts or GitHub issues is quite difficult when a major part of the code is not provided in the post or issue. In the end, we have selected a set of 11 real-world faulty DRL programs that contain faults related to 8 of 11 types of faults in the taxonomy. Table \ref{resultsTable} summarizes attributes of these real faulty DRL programs. Figure \ref{fig:snippet2} shows a faulty DRL program extracted from SO post \#50308750 (program No. 7 in Table \ref{resultsTable}). In this example, the developer failed to properly close the environment at the end of the program (indicated with 1 in the code snippet). This fault belongs to Type 3, so the correct execution of the program depends on the successful termination of the current episode while the symptom is expressed as a runtime error.\\
We have prepared synthetic buggy DRL programs by injecting faults into a clean DRL code as follows. First, one of the authors developed a clean DRL program using OpenAI Gym and TensorFlow which are two popular frameworks for DRL. We have implemented the DQN algorithm to solve the well-known CartPole-v0 problem \cite{cartpole} following two official tutorials \cite{tensordqn,pytorchdqn}. The code has been executed making sure that it can solve the problem properly by achieving a reasonable reward over 50 consecutive trials. Then, another author (different from the person who developed the clean code) injected faults into the clean DRL code. We followed fault patterns observed in real-world samples to inject bugs and generate synthetic faulty samples. Hence, the injected bugs are realistic reproductions of faults. For each type of fault, at least one faulty example is constructed. We have constructed multiple faulty examples when there is more than one context in which the detection rule can be triggered. For instance, let’s consider Type 5: \textit{Suboptimal  exploration  rate}. A suboptimal exploration rate may occur due to a wrong/missing update equation of exploration rate, suboptimal value of exploration rate, or suboptimal value of decay factor. Since we have collected some real-world DRL faulty programs, those examples have been manually analyzed to realize various contexts of faults. For more details, please see the source code of \tool{} containing all buggy examples \cite{drlinter}. Finally, we have ended up with a total of 15 faulty synthetic programs.\\
While we have used all samples in Table 3 for preparing synthetic faulty programs, to select some of them for direct evaluation by \tool{}, first we have excluded those developed by libraries other than Tensorflow and OpenAI Gym which are out of scope of the current version of \tool{}. In the next round, programs written with older versions of libraries were discarded if the API related to the fault was not supported in later versions. Finally, we ended up with 6 buggy DRL programs, namely No. 3 to 8 in Table 3.

\newcounter{rownumbers}
\newcommand\rownumber{\stepcounter{rownumbers}\arabic{rownumbers}}
\begin{table*}
\caption{Samples of real DRL faulty programs selected from SO and GitHub to reproduce synthetic buggy samples for evaluating \tool{}.}
\begin{center}
\resizebox{\textwidth}{!}{
\begin{tabular}{|c|c|c|c|c|}
 \hline
 \textbf{No.} & \textbf{SO\#}(\textbf{URL}) & \textbf{Symptom} & \textbf{Recommended Fix} & \textbf{Fault Type}\\
 \hline
 \hline
 \rownumber & 57106676 & Unstable learning, & Increase the update frequency of the target network & Type 7\\
  & & increasing loss & &\\
 \hline
 \rownumber & 56964657 & Unstable learning, & Increase the update frequency of the target network & Type 7\\
   & & increasing loss & &\\
 \hline
 \rownumber & 47750291 & Bad performance & Use an exploration mechanism & Type 4 \\
 \hline
 \rownumber & 54385568 & Bad performance & Decrease the exploration rate & Type 5\\
 \hline
 \rownumber & 51425688 & Bad performance & Decrease the exploration rate & Type 5\\
 \hline
 \rownumber & 49035549 & Bad performance & Decrease the exploration rate, Improve DNN design & Type 5\\
 \hline
 \rownumber & 50308750 & Compile-time error & Add proper API to close environment & Type 3\\
 \hline
 \rownumber & 47643678 & Bad performance & Detect the terminal state properly & Type 2\\
 \hline
 \rownumber & 40896951 & Bad performance & Change the activation of the last layer & Type 11\\
 \hline
 \rownumber & 37524472 & Bad performance & Change Q-learning update equation & Type 6\\
 \hline
 \rownumber & URL$^{\mathrm{1}}$ & Compile-time error & Detect state and action correctly & Type 8\\
 \hline
 \multicolumn{4}{l}{$^{\mathrm{1}}$https://github.com/tensorforce/tensorforce/issues/697.}
\end{tabular}}
\end{center}
\label{resultsTable}
\end{table*}

\subsection{Results}
\tool{} is first evaluated on 15 synthetic samples to investigate the correctness and preliminary effectiveness of the proposed approach. \tool{} has successfully detected the bugs in all synthetic examples. Our goal for using synthetic examples is mainly debugging, i.e., making sure of \tool{}’s accuracy and effectiveness prior to evaluating it on real-world examples. We have then tested \tool{} on 6 real-world faulty DRL programs. The results show that \tool{} can achieve a recall of 75\% and a precision of 100\% on these samples. We have evaluated the accuracy of \tool{} by checking the output of \tool{}, i.e., detected bug(s), and the bug(s) as they were acknowledged by SO users in the accepted answers. According to the recall value, \tool{} has not detected all existing faults in DRL programs. For example, program No. 6 in Table 3 suffers from two issues according to its accepted answer: suboptimal exploration rate and improper activation function in its DNN. While the former has been detected by \tool{}, the latter could not be detected since we have not considered the details of the employed DNNs in our meta-model of DRL programs. Moreover, the precision of 100\% means that while the tool may miss some faults in the evaluated DRL programs (overall recall is 75\%), it never detects bugs wrongly in the evaluated programs.\\
However, \tool{} is based on static analysis: it has been designed and implemented to detect faults that relate to structural (architectural) properties of DRL programs rather than their dynamic properties that need the programs to be executed. Based on the inspected faulty programs and the feedback we received in our survey, such faults are frequent in DRL programs and worth addressing. Indeed there are some bugs that could not be detected without dynamic analysis of the DRL programs, e.g., runtime bugs. In the future, we plan to expand our work through dynamic analysis to detect more faults (runtime faults and bugs in training of RL agents) and design issues (poor configuration or architectural choices) in DRL programs. So, lack of dynamic analysis is a limitation of the current version of \tool{} and is left for future work.\\
We have performed the experiments using a machine with Intel Core i5-3570 CPU and 16 GB of main memory running Windows 10. The average and standard deviation of \tool{} runtime for 30 runs per each faulty sample are 1.732 and 0.069 seconds, respectively. The graph checking (performed by GROOVE) consumes the main portion of the execution time, about 99.8\%, with an average of 1.728 seconds and standard deviation of 0.067. In fact, the graph extraction phase is accomplished by single or multiple passes through the code. Hence, the execution time grows linearly with the size of the DRL program (measured in terms of lines of code). On the other hand, GROOVE supports priority-based rule applications as well as various search strategies to explore the full state space, i.e., checking and applying all applicable rules in each state \cite{ghamarian2012modelling}. We have used BFS and priority-based rule application, to improve the efficiency of the graph checking. The running time of \tool{} can be improved by revising the implementation of rules in GROOVE which is left for future work.

\section{Threats to Validity}
\label{validity}
There are internal and external threats to the validity of this research. The taxonomy could be affected by biased when mining and labelling the artifacts. To address this issue, a clear systematic approach is followed in our study and mining/labeling is performed and evaluated by at least two of the authors through various rounds. We have extracted only "closed" issues from GitHub and with "at least one accepted" answer from SO ensuring that we analyzed only issues that were solved. Moreover, we conducted a survey to validate the proposed taxonomy with participants who have not been involved in the process of constructing it. Although we constructed the taxonomy using artifacts produced when developing four frameworks, we kept the title and description of fault types 
as general as possible. Therefore, popular frameworks were selected and we kept the categories as framework-independent as possible during labelling. Participants with different levels of expertise and background have validated the taxonomy as a means to address this threat. It is also possible that our questions and presented categories in the survey affected participant’s view directing them toward our proposed types of faults. To address this concern, we asked participants at the end of our survey to freely comment on our results and mention potential missing faults/issues in our study.\\
Basically, model-based static analysis of DRL programs should be helpful for practitioners/researchers to find bugs with this kind of low-cost analysis. However, the lack of behavioral analysis of DRL programs is a limitation of our approach, more faults could be discovered by such analysis. Some other bugs in DRL programs need further analysis, e.g. \textit{Wrong calculation of gradients} which is not covered by \tool{}. Moreover, limited access to real DRL programs along with verified bugs and recommended fixes to evaluate DRL testing approaches accurately and effectively could be regarded as a barrier in this line of research. It should be noted that in \tool{} faulty DRL programs are converted to model (graph) and then fault detection is performed at the level of model not code. In this way, the bias toward making the faults detectable by \tool{} is decreased. Since the proposed fault detection approach is model-based, the process of extracting models from DRL programs should be improved to increase the applicability of \tool{}.
\section{Related Work}
\label{relatedwork}
Although the number of studies on RL, DRL and DRL-based software systems has increased dramatically over the past decade, far less research has focused on testing RL systems. To the best of our knowledge, Trujillo et al. \cite{trujillo2020} is the very first work in this field. They have used neuron coverage as a well-known whitebox testing technique to investigate the evolutionary coverage of deep networks for the specific case of DRL \cite{trujillo2020}. 
Two different models of DQN have been tested for the Mountain Car Problem (MCP). Results revealed that observing good neuron coverage does not necessarily mean success in a RL task in terms of reward. A negative correlation is observed in their results and this is in contrast to typical DNNs confirming that neuron coverage is not capable of properly assessing the design or structure of DRL networks. The best possible coverage is achieved by extensive exploration which is not efficient in DRLs and does not help to maximize the reward.\\
There are some other research works that considered fault specifically in DL-based systems. A number of DL applications developed using TensorFlow have been studied by Zhang et al.  \cite{DL_bugs_1}. In their empirical study, they explored SO and Github to select faulty applications. From the selected faulty applications they obtained 175 TensorFlow related bugs. They manually examined the bugs 
to understand the challenges and strategies followed by developers to detect and localize the faults in the TensorFlow-based applications. They have presented some insights for two aspects of these faults: the root causes of bugs and the bugs impact on the application behaviour. Finally, authors classified the root causes and symptoms into seven and four different types, respectively. In this study, DRL applications that use some popular DRL frameworks, including OpenAI Gym, TensorForce, Dopamine and Keras-rl, have been studied. The popularity of these frameworks makes them representative of the state-of-the-practice in this field.
There is another methodological difference in the approach followed to mine 
SO posts: while Zhang et al. considered SO posts with at least one answer, we consider posts with an accepted answer. This additional condition is important to ensure 
that the studied faults were successfully solved and their solution was accepted by a developer. Moreover, their 175 collected bugs include generic faults, while our taxonomy covers only DRL specific faults. Finally, we did not restrict this study to the authors' analysis and have validated the presented taxonomy by conducting a survey with participation of DRL developers/researchers
.\\
Another characterisation of DL bug was reported by Islam et al. \cite{islam2019comprehensive}. Their study aimed to discover the most frequent types of bugs in DL programs as well as their causes and symptoms. A common pattern among bugs and its dynamics over the time has been studied as well. Similar to us, they have investigated a number of SO and GitHub bugs related to five DL frameworks: Theano, Cafe, Keras, TensorFlow, and PyTorch. They leveraged the list of root causes of bugs reported by Zhang et al. \cite{DL_bugs_1} to categorize the causes of bugs in their study. In contrast to our study, they analyzed various fault patterns and the correlation/distribution of bugs in different frameworks, and did not construct a taxonomy of faults.\\
The most recent and related research to this study is a validated taxonomy of real faults in DL systems presented in \cite{DLtaxo2020}. Humbatova et al. have constructed their taxonomy based on two sources of information: 1) manual analysis of Github artifacts and SO posts; and 2) interviews with developers/researchers. Their introduced taxonomy consists of 5 main categories containing 375 instances of 92 unique types of faults. The taxonomy is validated by conducting a survey with a distinct group of developers/researchers who verified the completeness and usefulness of the identified categories. 

\section{Conclusion}
\label{conclusion}
In this paper, a taxonomy of real faults in the DRL programs has been proposed. The methodology was manual analysis of faulty software artifacts from SO and GitHub developed using four selected DRL frameworks: OpenAI Gym, Dopamine, Keras-rl, Tensorforce. We manually analyzed artifacts and identified 761 issues. Then, we categorized the relevant issues through a multi-round labelling process. Finally, we obtained 11 distinctive types of faults that contain 27 faulty artifacts. The validation of the taxonomy has been performed by conducting a survey with 19 participants who have various backgrounds and levels of expertise in RL. The results have confirmed the relevance of identified types of faults in DRL programs.\\
Afterward, we have introduced \tool{}, a model-based fault detection approach for DRL programs by presenting a meta-model for these programs. The meta-model is represented by a type graph and graph transformation rules are used to implement the detection rules. A model of each DRL program is constructed by parsing its code to extract relevant information. Then, a graph checking process is performed to detect faults in the model. \tool{} has been evaluated successfully using synthetic and real faulty DRL programs. The results show that \tool{} can effectively detect faults with a recall of 75\% and a precision of 100\%. Currently, \tool{} could find faults in programs developed by TensorFlow and OpenAI Gym but the parser could be extended for other frameworks. Also, the proposed meta-model is designed for DQN-based DRL programs. However, it can be extended to support other approaches. Another direction of research for the future is expanding our set of fault detection rules to cover more types of faults and issues in DRL programs.

%
%

\begin{acknowledgements}
This work is partly funded by the Natural Sciences and Engineering Research Council of Canada (NSERC) and the Fonds de Recherche du Québec (FRQ).
\end{acknowledgements}

%
%


%
%
\bibliographystyle{spmpsci}      
\bibliography{biblio}

\end{document}